\pdfoutput=1

\documentclass[%
    aps                ,   
    prl                ,   
    reprint            ,   
    twocolumn          ,   
    showpacs           ,   
    groupedaddress ,   
    nofootinbib        ,   
    citeautoscript     ,   
    notitlepage        ,   
    floatfix
]{revtex4-1}
\usepackage[T1]{fontenc}
\usepackage[utf8]{inputenc}
\usepackage[english]{babel}
 
\addto\captionsenglish{\renewcommand{\figurename}{Fig.}}
\makeatletter
\renewcommand*{\fnum@figure}{{\normalfont\bfseries \figurename~\thefigure}}
\renewcommand*{\@caption@fignum@sep}{\textbf{ | }}
\makeatother
\usepackage{amsmath}
\usepackage{amssymb}
\usepackage{accents}

\usepackage{nicefrac} 
\usepackage{mathrsfs}
\usepackage{soul}
\usepackage{mathtools}
\usepackage{dsfont}

\usepackage[hyperref]{xcolor}
\usepackage{graphicx}
\usepackage{adjustbox}
\graphicspath{{figures/PNG/}{figures/PDF/}{figures/}}
\usepackage{quantikz}

\usepackage[
    unicode           = true  ,
    plainpages        = false , 
    pdfpagelabels     = true  , 
    bookmarks         = true  ,
    bookmarksnumbered = true  ,
    bookmarksopen     = true  ,
    breaklinks        = true  ,
    backref           = false ,
    colorlinks        = true  ,
    linkcolor         = blue  ,
    urlcolor          = blue  ,
    citecolor         = red   ,
    anchorcolor       = green ,
    hyperindex        = true  ,
    linktocpage       = true  ,
    hyperfigures      = true
]{hyperref}
\hypersetup{
    linkcolor         = [RGB]{000 114 189} ,  
    urlcolor          = [RGB]{000 114 189} ,  
    filecolor         = [RGB]{000 114 189} ,  
    citecolor         = [RGB]{217 083 025} ,  
    anchorcolor       = [RGB]{119 171 048} ,  
    pdftitle={},
    pdfauthor={}
}
\usepackage{cleveref}
\usepackage{orcidlink}

\usepackage[normalem]{ulem}

\begin{document}

\title{The true cost of factoring: Linking magic and number-theoretic complexity in Shor's algorithm}

\author{Alessio Paviglianiti\,\orcidlink{0000-0002-2719-7080}}
\email[]{alessio.paviglianiti@epfl.ch}

\author{Matteo Seclì\,\orcidlink{0000-0002-9608-096X}}

\author{Emanuele Tirrito\,\orcidlink{0000-0001-7067-1203}}

\author{Vincenzo Savona}

\affiliation{Institute of Physics, École Polytechnique Fédérale de Lausanne (EPFL), CH-1015 Lausanne, Switzerland \\
Center for Quantum Science and Engineering, Ecole Polytechnique Fédérale de Lausanne (EPFL), CH-1015 Lausanne, Switzerland}

\date{\today}

\begin{abstract}

The execution cost of quantum algorithms is typically quantified through asymptotic gate counts and qubit register sizes, yet these metrics do not directly capture which genuinely quantum resources, and in what amount, must be created and maintained for the computation to succeed. The systematic quantification of such information-theoretic requirements in quantum computing protocols remains an extremely challenging open problem, despite their direct role in establishing quantum advantage. To address this gap, we investigate the generation of non-stabilizerness (or magic), one of the key resources, in the paradigmatic Shor's factoring algorithm, revealing a deep connection between intrinsic quantum complexity and the computational hardness of the underlying number-theoretic problem. By developing an explicit analytic theory, we demonstrate the fundamental role of magic in the successful execution of the algorithm, and show that Shor's routine maximally exploits the quantum resource in practically relevant regimes. Our findings create a concise conceptual link between the classical algorithmic difficulty of a task and the non-stabilizer price to solve it on quantum hardware, complementing standard circuit-cost analyses with a resource-based metric that is naturally aligned with the real bottlenecks of fault-tolerant quantum computing.

\end{abstract}

\maketitle

\section{Introduction}

Quantum devices are anticipated to revolutionize modern technology by solving classically intractable problems, with prospective groundbreaking applications spanning engineering, physics, chemistry, and life sciences. The fundamental ingredient for quantum advantage is the exploitation of quantum resources, which are intrinsically difficult to simulate on classical machines~\cite{daley2022,bravyi2018}. As a consequence, understanding the essential quantum-information requirements for the successful execution of quantum protocols is crucial for developing effective physical implementations. Despite the critical importance of these quantum resources, a systematic mapping of their generation across quantum algorithms is still lacking, as the computational cost of quantum computing is more commonly quantified through gate and qubit counts. As such, the development of a resource-theoretic characterization of cost stands as a major open frontier in quantum information theory.

Among these key resources, \textit{non-stabilizerness}, or \textit{magic}~\cite{bravyi2005,bravyi2012}, emerges as a fundamental metric of quantum complexity, quantifying the difficulty of simulating a state classically through the well-established stabilizer formalism~\cite{gottesman1998,aaronson2004}. While the stabilizer framework is extremely powerful for investigating many-body physics and forms the pillar of quantum error correction~\cite{knill1997,Veitch_2014}, non-stabilizerness is a necessary ingredient to achieve universal quantum computation and unlock truly quantum behavior~\cite{bravyi2019}. From this perspective, magic not only separates easy-to-simulate states from genuinely complex ones, but also estimates how many non-stabilizer operations are needed to prepare them~\cite{eastin2009,capecci2025}.

\begin{figure}[t!]
    \centering
    \includegraphics[width=0.9\columnwidth]{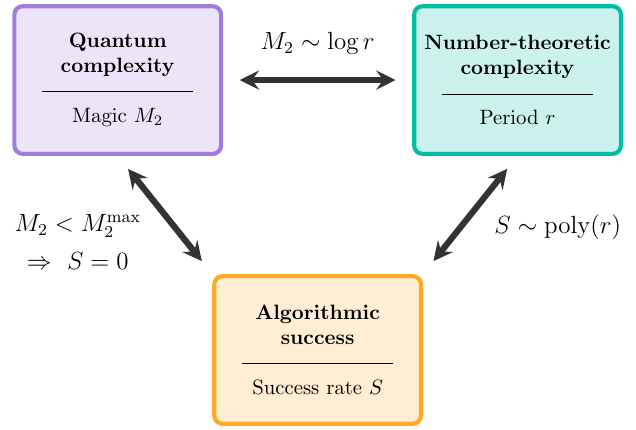}
    \caption{Sketch representation of the main findings of this work. Quantum complexity, quantified by magic, scales with the number-theoretic difficulty of determining the period $r$ in the period-finding task performed by Shor's circuit. Moreover, algorithmic success correlates with computational hardness and requires the preservation of non-stabilizerness, as any loss of quantum information leads to immediate failure.}    
    \label{f:results_summary}
\end{figure}

In this work, we address the challenge of quantifying resources in quantum protocols by investigating magic in Shor's factoring algorithm, the paradigmatic example of super-polynomial quantum-over-classical advantage~\cite{shor1997,nielsen2010,mermin2007,smolin2013,peter2025}. Combining numerical simulations and a specifically developed analytic theory, we fully characterize the behavior of non-stabilizerness throughout the quantum circuit. Most importantly, we explore the connection between quantum complexity, number-theoretic hardness of the computational problem, and algorithmic success rate, revealing that these apparently distinct concepts are deeply intertwined (see Fig.~\ref{f:results_summary}).

\begin{figure*}[ht!]
    \centering
    \includegraphics[scale=0.291]{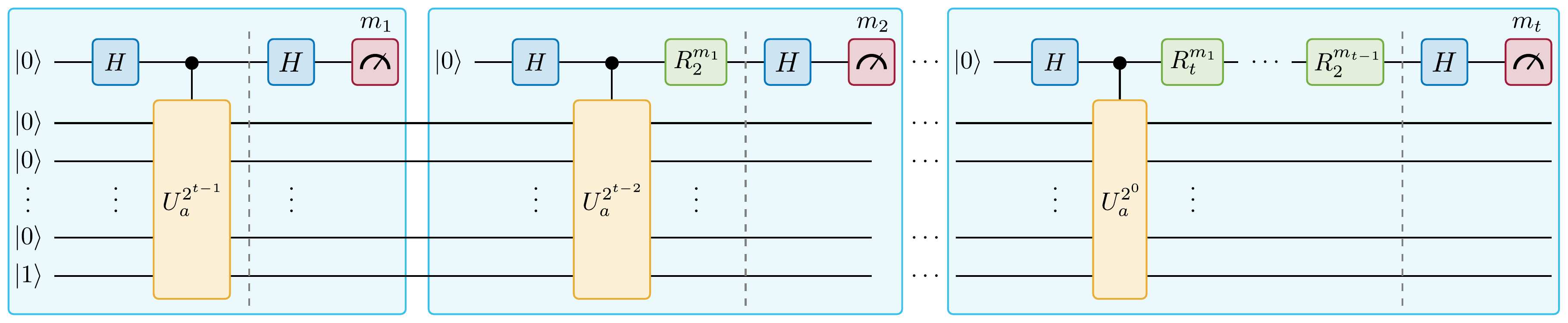}
    \caption{Quantum circuit for Shor's factoring algorithm. Each step (cyan boxes) implements a regular sequence of gates followed by a measurement of the QFT qubit. The outcomes $m_\tau$ determine the rotations implemented by the phase gates $R_k = \exp\left(i \pi Z/2^{k-1}\right)$ (green unitaries). In our analysis, magic is evaluated at the positions marked by the dashed vertical lines, which is completely equivalent to computing it right before the projective measurements.}
    \label{f:circuit}
\end{figure*}

As our main result, we demonstrate a strong correlation between the magic generated by the circuit and the intrinsic classical difficulty of the computational task. This reinforces the understanding that complex problems inherently require more quantum resources to be solved on quantum hardware. We further consolidate this notion by showing that any loss of non-stabilizerness directly causes the failure of the algorithm, emphasizing the importance of preserving quantum information for its success. Additionally, our findings clarify that large-scale implementations of Shor's algorithm, which constitute the regime of practical interest for applications, will consistently generate near-maximal quantum complexity, thereby fully exploiting this resource. Our analysis highlights the essential resource-based cost of practical implementations, as supplying non-Clifford operations via magic states represents the dominant overhead in leading fault-tolerant schemes~\cite{litinski2019,beverland2021,gidney2024}.

\section{Shor's Algorithm}

Shor's factoring algorithm is a paradigmatic protocol in quantum computing, providing a striking example where quantum-over-classical advantage is super-polynomial in the size of the task. The algorithm searches for a prime factor $p$ of an integer $N$. This is achieved by randomly choosing an integer $a<N$ which is coprime with $N$, i.e., $\mathrm{gcd}(a,N)=1$. Each coprime $a$ defines a modular exponentiation function $f(x)=a^x\,\,\mathrm{mod}\,N$, that is periodic with a period $r$, also referred to as the order of $f$, defined as the smallest nonzero integer such that $f(r)=1$. Shor's algorithm then returns an estimate of an integer multiple of $1/r$, from which the order $r$ can be recovered with high probability using a continued fraction expansion~\cite{nielsen2010,mermin2007}. If the period is even, its knowledge enables the factorization of $N$, solving the task with an overall cost of $\mathcal{O}(\mathrm{poly}(\mathrm{log}\,N))$ operations.

Different coprimes $a$ can feature modular exponentiation functions with distinct periods $r$. If $r$ is small, a brute-force search allows to determine it quickly, making the problem easy to solve. In contrast, large values of $r$ are notoriously difficult to extract, requiring a classical computational cost of $\mathcal{O}(\sqrt{r})$~\cite{pollard1978}. This suggests that $r$ can be used as a metric of the \textit{number-theoretic complexity} of the task. In general, coprimes with small $r$ are rare, and most coprimes are associated with periods $r\sim\mathcal{O}(N)$~\cite{luca2002}.

In this work, we consider the implementation of Shor's algorithm shown in Fig.~\ref{f:circuit}~\cite{beauregard2003,mosca1999}. The circuit realizes a $t$-step semi-classical quantum Fourier transform (QFT)~\cite{griffiths1996}, where a single QFT qubit is sequentially measured to output a bitstring $\mathbf{m} = (m_1,\dots, m_t)$ of measurement outcomes. The measurements are interleaved with controlled modular multiplications $C-U_a^{2^k}$, where $U_a\ket{y}=\ket{a\cdot y\,\mathrm{mod}\,N}$ acts on an $n$-qubit register with $n = \lceil \log_2 N \rceil$. These unitaries encode the information on the period $r$. Finally, the QFT outcomes $\mathbf{m}$ provide an approximation of the quotient $s/r$, where $1\leq s<r$ is an unknown integer. The number of algorithm steps $t$ directly controls the accuracy of the QFT estimate, and a common safe choice to ensure sufficient precision is $t = 2n + 1$~\cite{nielsen2010}.

\ 
\section{Magic and Quantum Complexity}

A key concept in quantum information theory is the definition of \textit{quantum complexity}. This notion characterizes the difficulty in simulating a given quantum state using purely classical resources, and thus provides a criterion to identify genuinely quantum behavior. While a unique and comprehensive definition of quantum complexity is missing, as quantum states can be complex in multiple ways, measures like entanglement or non-Gaussianity characterize computational hardness with respect to specific classes of classically treatable states.

We focus our investigation on \textit{magic}~\cite{bravyi2005,bravyi2012}, a measure of complexity based on the stabilizer formalism. Stabilizer states, generated by the Clifford operations $\{\mathrm{CNOT},H,S\}$, constitute a family that can be simulated efficiently on classical computers~\cite{gottesman1998_2} with $\mathcal{O}(\mathrm{poly)}(L)$ resources, where $L$ is the number of qubits and $L=n+1$ for the case studied here. Despite their accessibility, stabilizer states can exhibit quantum features like maximal entanglement, and have been successfully applied to describe genuine many-body effects such as correlation spreading~\cite{nahum2017}, dynamical phase transitions~\cite{li2019}, and topological order~\cite{kitaev2003}. Importantly, the set of Clifford gates is not universal for quantum computation. Generating arbitrary states requires non-Clifford operations, the most common choice being the $T$-gate $T = \mathrm{diag}(1,e^{i\pi/4})$~\cite{bravyi2016}. The application of $T$-gates generally leads a state outside the stabilizer set and increases the classical cost of simulating it~\cite{bravyi2016,bravyi2019}. Within this framework, magic provides a measure of how distant the state is from the efficiently-simulable class, and thus quantifies the amount of quantum resources needed to realize it.

Non-stabilizerness can be quantified through the Stabilizer Rényi Entropy~\cite{leone2022} (SRE)
\begin{equation}\label{SRE}
    M_2(\ket{\psi}) = -\log \left(\sum_{P \in \mathcal{P}_L} \frac{\bra{\psi}P\ket{\psi}^4}{2^L}\right),
\end{equation}
where $\mathcal{P}_L = \{I,X,Y,Z\}^{\otimes L}$ is the set of $L$-qubit Pauli strings. This measure vanishes if and only if $\ket{\psi}$ is a stabilizer state, signaling zero complexity, and is upper bounded by $L \log 2$. Moreover, evolving a state with a Clifford unitary leaves $M_2$ unchanged, as this operation does not increase the complexity of simulation with the stabilizer formalism. Finally, the SRE behaves additively for product states. These properties, combined, make Eq.~\eqref{SRE} a proper measure of quantum complexity from the perspective of quantum resource theory. Since its introduction, the SRE has been applied to a wide range of many-body systems and circuits, linking it to quantum chaos~\cite{leone2021} and criticality~\cite{haug2023,tarabunga2023,tarabunga2024,lami2023}. Moreover, notable progress has been made in estimating SRE in experimental settings~\cite{oliviero2022,haug2023_2,bluvstein2024logical,niroula2024}.

\

\section{Evolution of Magic in Shor's Algorithm}
\begin{figure*}[t!]
    \centering
    \includegraphics[width=\textwidth]{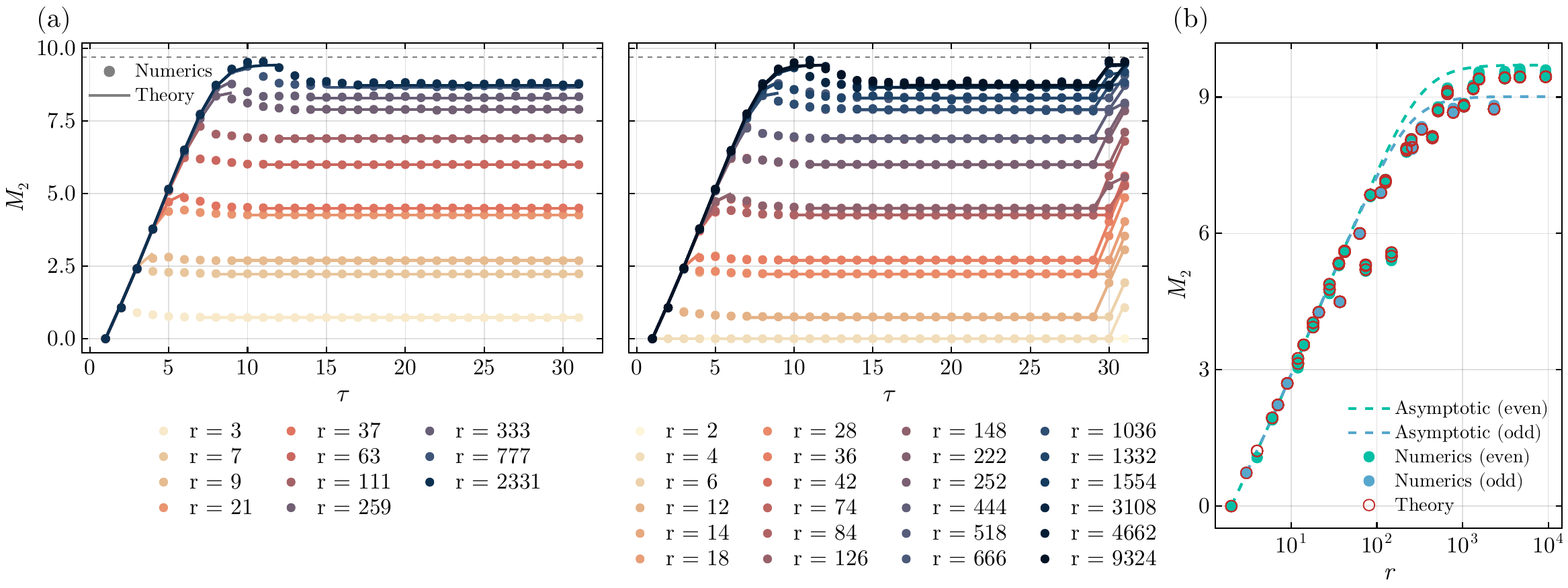}
    \caption{Magic in Shor's algorithm, using $N = 18923$. (a) Evolution of magic as a function of the discrete time $\tau$, for odd (left) and even (right) values of the period $r$. For each $r$, a random coprime $a$ with that $r$ is used. The theoretical prediction (solid lines) is obtained from Eq.~\eqref{magic_simple_formula}. The dashed line shows the average magic of Haar-random states for comparison. (b) Scaling of the final magic $M_2(\tau=t)$ as a function of $r$. The asymptotic curves are evaluated using Eq.~\eqref{closed_Lambda}. For each $r$, $10$ different random choices of $a$ are shown (if available). For all plots,
    numerical results are averaged over $150$ random realizations of the algorithm.}
    \label{f:magic}
\end{figure*}

The driving question of our investigation is to understand whether the number-theoretic complexity of order-finding, quantified by the period $r$ itself, is linked to the quantum complexity of the states generated by Shor's algorithm. To this purpose, we evaluate the SRE throughout the steps of Shor's algorithm, considering several coprimes associated to distinct periods. Beyond numerical simulations, we develop an analytic theory of the dynamics of magic, i.e., the evolution of the SRE of the quantum state along the circuit. As our main finding, we observe a striking correlation between magic and number-theoretic complexity, indicating that larger values of $r$ require a larger amount of quantum resource, eventually approaching its theoretical upper bound.

We apply Shor's algorithm to the factorization of $N = 18923 = 127\cdot 149$. For this choice, $n = 15$ qubits are required to store integers modulo $N$, and $L = 16$ qubits are used in total. The coprimes of $N$ feature a total of $35$ distinct periods $r$ of the modular exponentiation functions. We set the number of QFT steps to $t=2n+1$, which provides sufficient accuracy for the estimation of $1/r$~\cite{nielsen2010}. At every step, which we label by $\tau \in [1,t]$ and can be regarded as a discrete time of the algorithm, magic is evaluated before the Hadamard gate preceding each measurement, as shown in Fig.~\ref{f:circuit}. Since $H$ gates do not change the SRE, this is completely equivalent to computing it after the Hadamards, but this choice is more convenient for the analytic calculation of magic discussed below.

The dynamics of the SRE exhibits a particular and detailed structure, in sharp contrast with the cases of generic random circuits~\cite{paviglianiti2025,turkeshi2025,magni2025quantum} or Hamiltonian evolution~\cite{odavic2025,tirrito2025_anti,tirrito2025}. As shown in Fig.~\ref{f:magic}(a), we observe three key regimes with extents that strongly depend on $r$. In the following, we assume that the period factorizes as $r=2^k \Tilde{r}_\mathrm{odd}$, where $k\geq 0$ is an integer and $\Tilde{r}_\mathrm{odd}$ is an odd number. At early times, $M_2$ features an approximately linear ramp, ending at a $\tau^* = \lceil \log_2 \Tilde{r}_\mathrm{odd}\rceil$. Then, the non-stabilizerness relaxes to a plateau value, whose magnitude is also visibly correlated with the period. Finally, only if $r$ is even, the SRE exhibits a final growth in the last $k$ steps. From these results, we immediately notice a clear correlation between $r$ and the amount of magic generated by the algorithm. In particular, $M_2$ approaches the average value of Haar-random states~\cite{turkeshi2025} for the largest values of $r$, signaling that the required non-stabilizerness is close to maximal.

All three time regions are captured theoretically by the analytic model we developed. Upon careful inspection, we observe that the state at most steps of the algorithm features a highly structured superposition, where different computational basis states appear with uniform weight and pseudo-random phases as
\begin{equation}\label{superposition}
    \ket{\psi} = \sum_{\mathbf{m}\in\mathcal{D}}\frac{e^{i\theta_\mathbf{m}}}{\sqrt{D}}\ket{\mathbf{m}}.
\end{equation}
Here $\mathcal{D}$ is the set of bitstrings appearing in the superposition with non-zero amplitudes, and $D=\dim \mathcal{D}$. This key observation, derived and discussed in great detail in the Supplemental Material, enables the analytic characterization of magic, yielding the closed formula
\begin{equation}\label{magic_simple_formula}
    M_2 = 4 \log D - \log\left( 4\Lambda + 6 D^2 - 5 D\right).
\end{equation}
Here $\Lambda = \sum_{\mathbf{m}\neq\mathbf{n}\neq\mathbf{p}\neq\mathbf{q}\in \mathcal{D}} \delta_{\mathbf{m}\oplus\mathbf{n}\oplus\mathbf{p}\oplus\mathbf{q},\mathbf{0}}$ is a geometric contribution that depends on the specific elements of $\mathcal{D}$, which are the bitstrings generated by the gates $C-U_a^{2^k}$.

Throughout the initial ramp $\tau \leq \tau^*$, the number of states $D$ in superposition doubles at each step, explaining the linear increase of the SRE. At the start of the plateau region $\tau \gtrsim \tau^*+1$, Eq.~\eqref{superposition} no longer holds, and the state can exhibit a non-uniform superposition. Nevertheless, for $\tau \gg \tau^*$ the state recovers a highly structured superposition: the QFT qubit disentangles from the rest of the system and becomes a stabilizer, whereas the remaining $n$-qubit register is well-described by Eq.~\eqref{superposition} once again, with $D=\Tilde{r}_\mathrm{odd}$. This fully explains the numerical evidence of Fig.~\ref{f:magic}(a): after reaching the peak value, the SRE relaxes to a value smaller by $\mathcal{O}(1)$ because the QFT qubit no longer contains any non-stabilizerness.
Finally, if $r$ is even, the last $k$ steps feature the same behavior as the early time regime, with $D$ doubling at each step and Eq.~\eqref{superposition} holding once again. As a consequence, $M_2$ restarts its growth, in agreement with the numerical results.

For $\tau=1$, we observe that $M_2$ vanishes. Indeed, it can be shown exactly that when $D=2$ the state is always a stabilizer. This is particularly relevant for the special case of $r=2$, for which the state remains a stabilizer throughout the full algorithm.

\section{Correlating Quantum and Number-Theoretic Complexity}

The magic developed by the circuit shows a strong correlation with the magnitude of the period $r$. This connection is further highlighted in Fig.~\ref{f:magic}(b), which displays $M_2$ evaluated at the last step $\tau=t$ for different choices of $r$. The quantum complexity grows linearly with $\log r$, until it eventually saturates for $r \gg 2^{L/2}$. For finite system size $L$, the SRE is not fully determined by $r$ alone, as it preserves a weak dependence on the choice of the coprime $a$, encoded in the geometric term $\Lambda$ appearing in Eq.~\eqref{magic_simple_formula}. Nevertheless, when $L\gg 1$ the bitstrings contained in the superposition set $\mathcal{D}$ behave as if randomly distributed, and $\Lambda$ can be estimated from a combinatorial argument. As the key result of our work, in this asymptotic limit magic is only a function of the period, thus establishing a direct connection between the number-theoretic difficulty of order-finding and the quantum resources needed to run Shor's algorithm. 

Assuming $L\gg 1$, we can capture analytically the asymptotic dependence of $M_2$ on $r$. At small periods $r \ll 2^{L/2}$, the term $\Lambda$ is negligible, as the size of $\mathcal{D}$ is relatively small compared to the set of all possible bitstrings. As a consequence, the SRE is approximately given by $M_2(r) \approx \log\left(\frac{r^3}{6r-5}\right)$. In the opposite limit of $r \gg 2^{L/2}$, instead, the saturation value is equal to $M_2 \approx (L-2-\epsilon)\log 2 -3\cdot 2^{L-\epsilon-1}/r^2$ at leading order, where $\epsilon=0$ if $r$ is even and $\epsilon = 1$ if it is odd. We remark that the leading order $L \log 2$ is the theoretical upper bound to the SRE, signaling maximal quantum complexity. 

\section{Period Occurrence and Success Rate}
\begin{figure}[ht!]
    \centering
    \includegraphics[width=\columnwidth]{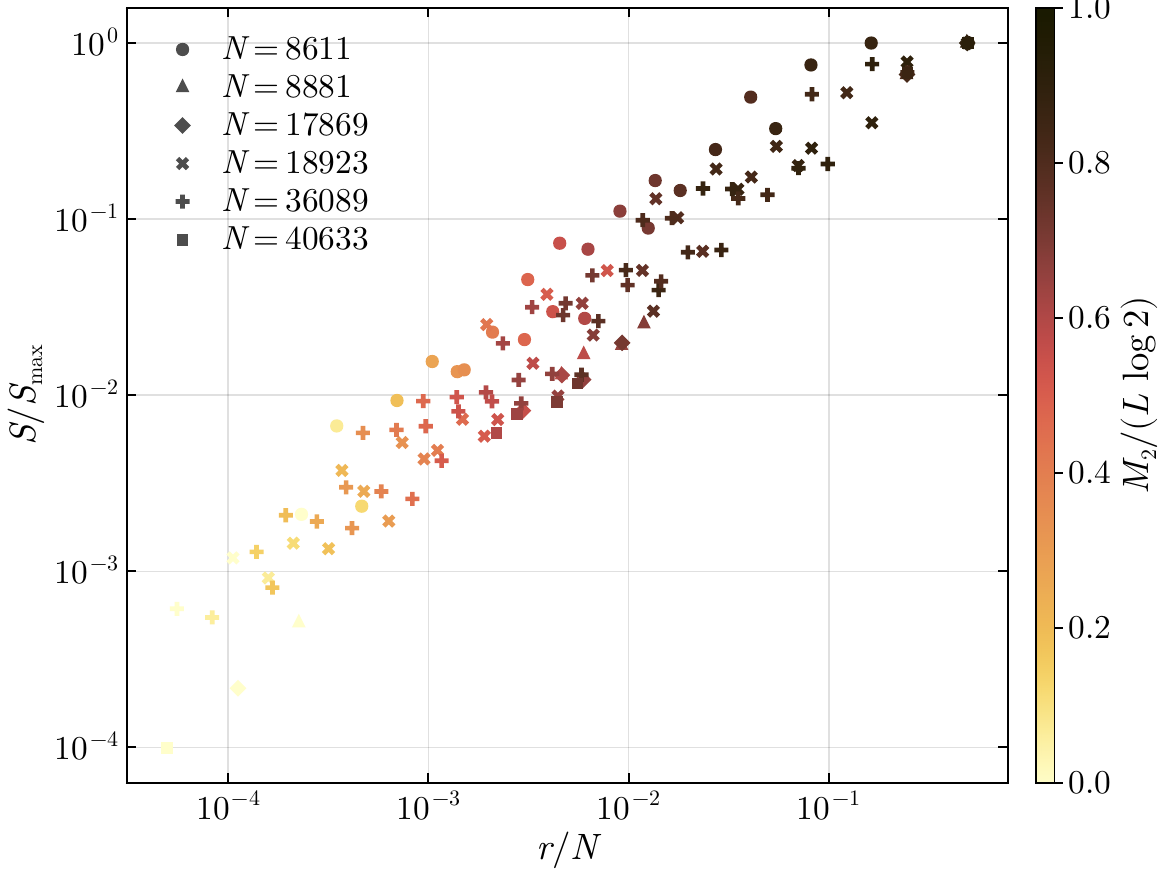}
    \caption{Conditional algorithmic success rate $S/S_\mathrm{max}$, normalized by its maximal value, as a function of the rescaled period $r/N$. The SRE density $M_2/(L \log 2)$ is visualized through colors. The main contribution to the global algorithmic success rate comes from large periods, which require close-to-maximal magic to be simulated. We present data for six choices of products of primes $N$, covering the range $L\in{[15,17]}$. The success rate is normalized by its maximum to account for different number of periods for different $N$. For each $r$, $M_2$ and $p_\mathrm{succ}$ are averaged over $100$ different coprimes $a$ (if available). $p_\mathrm{succ}$ is further averaged over $100$ independent realizations of the algorithm for each $a$. The SRE is evaluated analytically using Eq.~\eqref{magic_simple_formula}.}
    \label{f:success_rate}
\end{figure}

The above analysis highlights the existence of small periods that exhibit modest generation of SRE. A natural question is then whether these can be used to solve the factoring problem instead of the larger ones, which bear higher quantum complexity. It is known, however, that small values of $r$ are extremely rare for $N\gg1$~\cite{luca2002}, which implies that the generation of large magic is inevitable in practice.

The overall probability of successful detection of a specific period $r$ is determined by two factors. The algorithm operates by picking a random $a<N$, which will be a coprime of $N$ with high probability (otherwise, $N$ is easily factored). The first factor $g(r)$ is the occurrence frequency that $a$ has period $r$, satisfying $\sum_r g(r) = 1$. Then, once $a$ has been chosen, Shor's algorithm has a finite success probability $p_\mathrm{succ}$ of finding the period $r$ correctly through the continued fraction expansion.\footnote{Throughout our analysis, $p_\mathrm{succ}$ is evaluated using the textbook application of the continued fraction algorithm. More advanced postprocessing schemes have been developed~\cite{ekera2017,willisch2023}, and while they can increase the success probability, we do not expect they change our results qualitatively.} Combining these two factors, we introduce the conditional algorithmic success rate $S = g \cdot p_\mathrm{succ}$, which quantifies the actual contribution of a given period $r$ to the overall successful execution of the task. Notice that this is the probability of correctly finding a specific $r$, and not of actually factoring $N$, as additional conditions must be met for the latter.\footnote{Odd periods cannot be used to factor $N$. In addition, finding an even $r$ enables the factorization only if $a^{r/2} \neq -1 \,\,\mathrm{mod}\,N$.}

The conditional algorithmic success rate is presented in Fig.~\ref{f:success_rate}. We observe a clear correlation where $S$ appears to scale as $S \sim (r/N)^\alpha$ with $\alpha \approx 1$, indicating that large periods carry the vast majority of the overall probability. This behavior is mostly determined by the occurrence frequency $g$. Combining this with our previous result on quantum complexity, we conclude that the conditional algorithmic success rate is dominated by algorithmic realizations that produce large amounts of non-stabilizerness. In the limit of $L\gg 1$, we know analytically that the SRE approaches its maximum for $r \gtrsim 2^{L/2} \sim \sqrt{N}$. If we assume the relation $S \sim r/N$ supported by the above analysis, this implies that the region of Fig.~\ref{f:success_rate} with maximal magic extends to smaller values of $r/N$ as $N$ increases. As a consequence, the conditional algorithmic success rate of small periods becomes increasingly negligible, in line with the understanding that large semiprimes $N$ cannot be factored easily by relying on lucky cases with very small $r$. This evidence implies that, as $N$ grows, it becomes increasingly unlikely to solve the order-finding task by generating a limited amount of magic.

\section{Algorithmic Success Probability Under Loss of Quantum Information}

So far, we considered the magic produced when the algorithm depth $t$ is set to $2n+1$, which ensures sufficient accuracy for the estimate of $1/r$~\cite{nielsen2010}. We now consider the case where the number of steps is reduced, addressing the relevant question of whether the order-finding problem can still be solved using a shorter circuit~\cite{nam2020,coppersmith2002,oonishi2023}. Unsurprisingly, reducing $t$ causes a loss of success probability. However, we show that
$p_\mathrm{succ}$ does not drop to zero immediately for $t\lesssim2n+1$, but remains finite as long as the algorithm is still able to generate sufficient magic for its correct functioning. This observation establishes the loss of quantum information as the key mechanism underlying the failure of the algorithm.
\begin{figure}[ht!]
    \centering
    \includegraphics[width=\columnwidth]{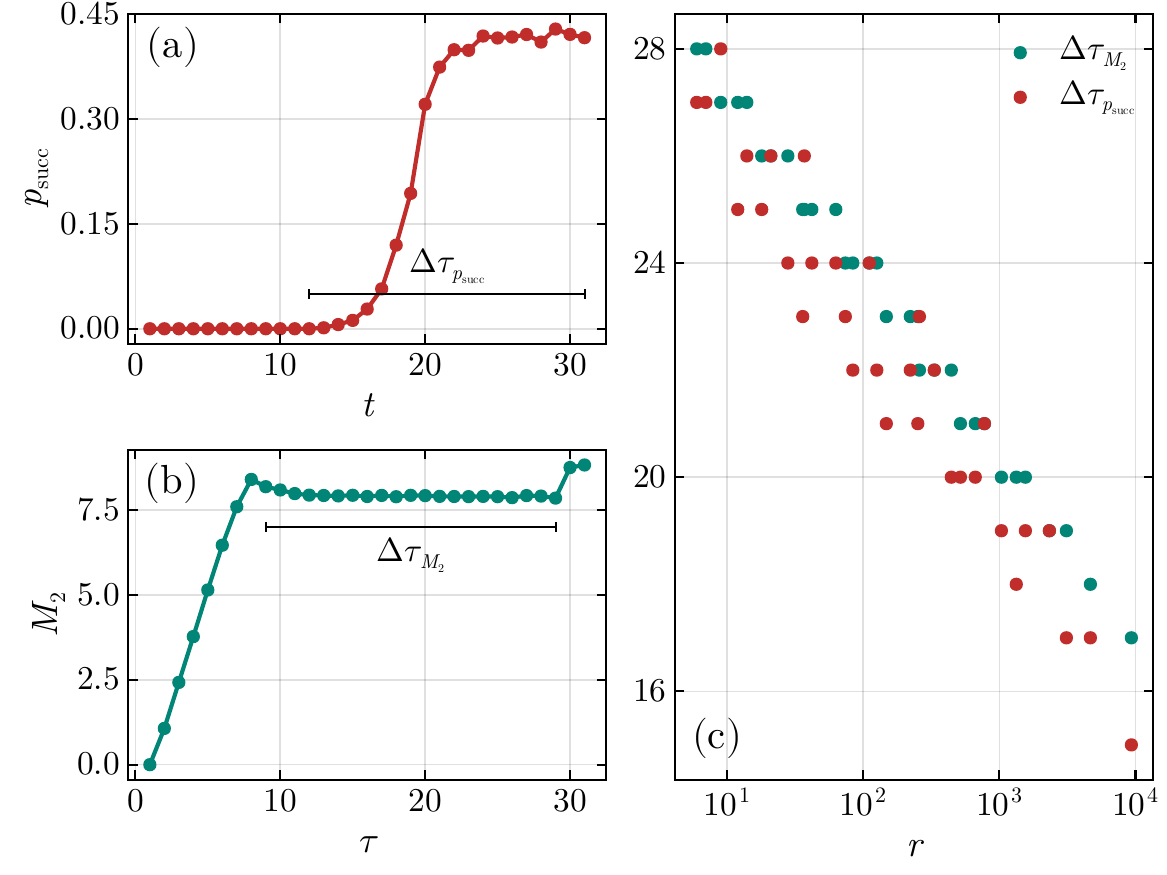}
    \caption{Magic plateau time $\Delta \tau_{M_2}$ compared to the decay time $\Delta \tau_{p_\mathrm{succ}}$ of the success probability, using $N=18923$. (a-b) Graphical visualization of the time intervals, using a random coprime $x$ with $r=1036$. (c) Agreement between the time intervals. The success probability is averaged over $5000$ random realizations of the algorithm. $\Delta \tau_{p_\mathrm{succ}}$ is estimated by looking for the value of $t$ where $p_\mathrm{succ}$ becomes zero within the sampling precision. } 
    \label{f:plateau}
\end{figure}

Our previous analysis of the SRE evolution clarifies that magic saturates to a plateau after an initial growth. From the analytic model, we know that $M_2$ stays flat up until $\tau = t-k$. Importantly, while the previous analysis assumed the specific value of $t=2n+1$, the same result applies to arbitrary algorithmic depth. If $t$ is increased, the SRE will still exhibit the same evolution shown in Fig.~\ref{f:magic}(a), just with a longer plateau. Similarly, reducing $t$ shrinks the flat region. Crucially, if the depth $t$ is reduced below $\lceil \log_2 r\rceil$, the plateau disappears completely, and the algorithm ends before magic reaches its maximal value shown in Fig.~\ref{f:magic}(b).

A natural question is then to clarify whether this loss of non-stabilizerness is related to the success rate of the algorithm. To investigate this, we evaluate the success probability $p_\mathrm{succ}$ for $t\leq t_\mathrm{max} = 2n+1$, and define the $\Delta \tau_{p_\mathrm{succ}}$ as the time interval for which it remains finite (see Fig.~\ref{f:plateau}(a)). In parallel, we introduce the magic plateau length $\Delta \tau_{M_2} = t_\mathrm{max} - \lceil \log r\rceil$, which is known from the theoretical analysis and is illustrated in Fig.~\ref{f:plateau}(b). As shown in  Fig.~\ref{f:plateau}(c), these two time intervals are highly correlated, suggesting that Shor's algorithm remains functional as long as sufficient quantum complexity is maintained.

These results establish a connection between non-stabilizerness and algorithmic success, consolidating magic as a fundamental and necessary resource for the execution of the order-finding task. In addition, $\Delta\tau_{M_2}\approx \Delta \tau_{p_\mathrm{succ}}$ determines how much the algorithm can be shortened while still being useful. We observe that larger periods require longer depths $t$, in agreement with the intuitive understanding that they are harder to estimate, as they are more complex from a number-theoretic perspective.

\section{Discussion}

The advantage of quantum algorithms over classical methods naturally requires quantum resources, such as magic or entanglement. We have studied the generation of non-stabilizerness in Shor's algorithm, highlighting a direct correlation between the quantum complexity produced by the circuit and the intrinsic computational difficulty of the order-finding problem. This result elevates magic to a powerful diagnostic tool for general computational complexity, extending its relevance well beyond its quantum information roots. Additionally, our analysis quantitatively establishes the non-stabilizerness requirements for the execution of the algorithm, thereby providing an essential lower bound on its cost in terms of magic resource states in practical fault-tolerant implementations~\cite{gidney2021,gidney2025,babbush2026,cain2026}.

By leveraging the special superposition structure of the state, we have developed an analytic theory of magic growth in Shor's algorithm. This enables a detailed and comprehensive understanding of its dynamics, and proves that the period $r$ encoded in the order-finding task bounds and controls the production of SRE. In particular, magic approaches its maximal value for large periods, which account for the vast majority of the total success probability of the protocol. 
Consequently, practical applications of the algorithm fully leverage this quantum resource, and realize states that can never be simulated on a classical computer at large system sizes.

Beyond these results, our work highlights the key role of preserving non-stabilizerness in the successful execution of the algorithm. We demonstrate that loss of magic causes the inevitable failure of order finding. This consolidates the familiar understanding that quantum complexity is a fundamental ingredient for the functioning of the algorithm, as it is responsible for quantum advantage. Crucially, this observation cautions against the risks of approximating the algorithm through gate truncations, which have been proposed previously in the literature~\cite{nam2020,coppersmith2002,oonishi2023}. Similarly, noise is expected to distort the complexity of the state, thus deteriorating the success rate by the same mechanism.

Our investigation paves the way for exploring the role of quantum magic in other quantum algorithms, which remains a compelling open question. While this work establishes a first link between non-stabilizerness and number-theoretic complexity, it will be crucial to test this connection across different problems, such as Grover's algorithm among others. In this direction, it would be interesting to study whether the behavior of magic changes in the presence of a polynomial quantum advantage, rather than an exponential one as in the case of Shor. Finally, while non-stabilizerness is a necessary ingredient to achieve genuine quantum complexity, other separate measures provide different perspectives on it. The systematic investigation of other quantum resources is key to developing a complete picture of how quantum and number-theoretic difficulty are correlated.

\section*{Methods}
\textbf{Analytic model} -- We hereby provide an analytic model for the SRE in Shor's algorithm. A complete derivation of this result is discussed in detail in the Supplemental Material.

The calculation relies on the formalism of the fast Walsh-Hadamard transform presented in Ref.~\cite{huang2026}, and exploits the special  superposition structure of Eq.~\eqref{superposition}. The SRE is given by
\begin{equation}
    M_2 = \begin{cases}
    0                                                   & \text{if} \,\,\, D=2,\\
    4 \log D - \log\left( 4\Lambda + 6 D^2 - 5 D\right) & \text{otherwise},
    \end{cases}
\end{equation}
where 
\begin{equation}\label{D_size}
    D = \begin{cases}
        2^\tau                 & \text{for} \,\,\,\tau\leq \tau^*,\\
        \Tilde{r}_\mathrm{odd} & \text{for} \,\,\,\tau^*<\tau\leq t-k,\\
        r/2^{t-\tau}           & \text{for}\,\,\,t-k<\tau.
    \end{cases}
\end{equation}
This result is quantitatively accurate for $\tau\leq \tau^*$ and $\tau-\tau^*\gg 1$. In the time regimes $\tau\leq \tau^*$ and $\tau>t-k$, in particular, Eq.~\eqref{superposition} models the true wavefunction accurately. For $\tau-\tau^* \gg 1$ with $\tau\leq t-k$, the QFT qubit disentangles from the other ones and is in a stabilizer state; hence, it does not contribute to the SRE. In this case, Eq.~\eqref{superposition} still correctly models the state of the $n$-qubit register, and the bitstring space dimension $D$ in Eq.~\eqref{D_size} refers to the $n$-qubit superposition only. For $\tau\gtrsim\tau^*+1$, instead, the state takes a more general form, and the analytic model does not provide a quantitatively reliable estimate of the SRE.

The geometric term $\Lambda$ is computed as follows. Let $A(\mathbf{x})$ be the number of unique distinct unordered pairs $\{\mathbf{m}, \mathbf{n}\} \subset \mathcal{D}$ such that $\mathbf{\mathbf{m}\oplus\mathbf{n}} = \mathbf{x}$. Straightforward combinatorics leads to $\Lambda = 4\sum_\mathbf{x} A(\mathbf{x})(A(\mathbf{x})-1)$, whose evaluation cost scales as $\mathcal{O}(D^2)$. The set $\mathcal{D}$ is obtained by applying the permutation unitaries $U_{a}^{2^{t-\tau}}$ to the initial $n$-qubit state $\ket{0}^{\otimes (n-1)} \ket{1}$. Importantly, for the purpose of our analysis, the gates $U_a^{2^{t-\tau}}$ are built from the knowledge of their associated permutation matrix, and not from elementary reversible operations, as this does not affect the calculation of magic.

In the asymptotic limit of $L\to\infty$, we can provide a closed-form approximation of $\Lambda$ based on the assumption that the bitstrings of $\mathcal{D}$ are random and uniformly distributed. This leads to
\begin{equation}\label{closed_Lambda}
    \Lambda = \begin{cases}
        \frac{D(D-1)(D-2)(D-3)}{2^{L-1}} \qquad \text{for}\,\,\,\tau^*<\tau\leq t-k,\\
        \frac{D(D-1)(D-2)(D-3)}{2^L}\qquad \text{otherwise}.
        \end{cases}
\end{equation}
The difference between the two cases is due to the number of qubits considered in the calculation of the SRE, which are $n = L-1$ for $\tau^*<\tau\leq t-k$ and $L$ otherwise. Eq.~\eqref{closed_Lambda} provides the asymptotic theoretical prediction presented in Fig.~\ref{f:magic}(b).

\textbf{Numerical Simulations} -- The state vector along the circuit shown in Fig.~\ref{f:circuit} is simulated exactly using sparse arrays for the unitary gates. The evaluation of the SRE is computationally demanding, as the cost of computing Eq.~\eqref{SRE} scales as $\mathcal{O}(4^L)$. We rely on the stochastic sampling approach developed in Ref.~\cite{lami2023}, which provides a direct, unbiased estimate of $M_2$ for matrix-product states (MPSs). In order to apply this technique, we first convert the state wavefunction into an exact MPS representation using standard methods~\cite{schollwock2011}, and we then follow the procedure described in Ref.~\cite{lami2023}. We stress that the conversion to MPSs is performed without truncation of the bond dimension, and thus no approximation is introduced. Finally, the SRE is estimated by sampling over $10^4$ Pauli strings.

The output of Shor's algorithm is the record of measurement outcomes $(m_1,\dots,m_t)$, which contains information on the period $r$. Specifically, the number $x = \sum_{j=1}^t 2^{j-t-1}m_j$ is, with high probability, a $t$-bit estimate of the fraction $s/r$, where $s$ is an integer in the interval $[0,r-1]$. To recover $r$, we generate the continued fraction expansion 
\begin{equation}
    x = \frac{1}{a_1 + \frac{1}{a_2 +\frac{1}{a_3 + \dots}}},
\end{equation}
which provides a sequence of rational approximations of $r$. The algorithm searches for a valid period among the denominators appearing in this sequence, and is considered as successful if $r$ is found. Importantly, if a submultiple of $r$ instead of the actual period is found, which can occur if $s$ and $r$ share a prime factor, the search is considered to have failed. The conditional success probability $p_\mathrm{succ}$ is evaluated by iterating this process over multiple random realizations of the circuit and performing an average.

\section*{Acknowledgments}

We acknowledge support by the Swiss National Science Foundation through Projects No.\@ 200020\_215172, 200021-227992, and 20QU-1\_215928, and as part of NCCR SPIN (grant number 225153).
M.S.\@ acknowledges funding from the Swiss Academy of Sciences (SCNAT) through the Swiss Quantum Initiative (SQI) Grant No.\@ 24\_1111.
E.T.\@ was funded by the Swiss National Science Foundation (SNSF) under Grant No.\@ TMPFP2\_234754. E.T.\@ acknowledges CINECA (Consorzio Interuniversitario per il Calcolo Automatico) award, under the ISCRA initiative and Leonardo early access program, for the availability of high-performance computing resources and support.

\clearpage                             
\vbox{}                             

\bibliographystyle{naturemagwithdoi}   
\typeout{}                              
\bibliography{references}

\renewcommand{\thesection}{S.\arabic{section}}
\renewcommand{\thesubsection}{\thesection.\arabic{subsection}}
\renewcommand{\thesubsubsection}{\thesubsection.\arabic{subsubsection}}
\renewcommand{\theequation}{S\arabic{equation}}
\renewcommand{\thefigure}{S\arabic{figure}}
\renewcommand{\thetable}{S\arabic{table}}
\makeatletter
\renewcommand{\p@subsection}{}
\renewcommand{\p@subsubsection}{}
\makeatother
\setcounter{equation}{0}
\setcounter{figure}{0}

\pagebreak
\widetext
\newpage
\begin{center}
\textbf{\large Supplemental Material for\texorpdfstring{\\}{} ``The true cost of factoring: Linking magic and number-theoretic complexity in Shor's algorithm''}
\end{center}

\section{Structure of the wavefunction}\label{s:wavefunction}
In this Section, we analyze the superposition structure of the states generated throughout the algorithm. As we will see, the state can be regarded as a uniform superposition of computational basis states, each carrying a phase that can be considered pseudo-random. This observation enables the analytic evaluation of non-stabilizerness.

\textbf{Short times} -- As presented in Fig.~\ref{f:circuit}, Shor's algorithm consists of a sequence of steps, each consisting of (i) a Hadamard gate on the QFT qubit, (ii) a controlled unitary gate that implements a permutation $\Pi^{2^{t-\tau}}$ (where $\tau=1,\dots, t$ indicizes the steps) of computational basis states in the $n$-qubit register, (iii) a phase gate, (iv) another Hadamard gate on the QFT gate, and (v) a projective measurement. Let $\ket{\psi_\mathrm{in}(\tau)}$ be the input state of step $\tau$, obtained right after the measurement of step $\tau-1$. The starting state of the circuit is given by $\ket{\psi_\mathrm{in}(0)}=\ket{0}\ket{1}$. In our analysis, magic is evaluated immediately before step (iv), and we refer to states at that stage as $\ket{\psi(\tau)}$. For the first values of $\tau$, it is easily observed that each step doubles the number of computational basis states in the superposition, generating the sequence
\begin{subequations}
    \begin{equation}\label{psi_1}
    \ket{\psi(1)} = \frac{1}{\sqrt{2}}\left(\ket{0}\ket{1}+\ket{1}\ket{\Pi^{2^{t-1}}(1)}\right),
\end{equation}
\begin{equation}
    \ket{\psi(2)} = \frac{1}{2}\left(\ket{0}\ket{1}+e^{i\pi m_1}\ket{0}\ket{\Pi^{2^{t-1}}(1)}+e^{i\varphi_2}\ket{1}\ket{\Pi^{2^{t-2}}(1)}+e^{i(\varphi_2 + \pi m_1)}\ket{0}\ket{\Pi^{2^{t-1}+2^{t-2}}(1)}\right)
\end{equation}
\begin{equation}\label{ramp_wavefunction}
    \ket{\psi(\tau)} = \frac{1}{2^{\tau/2}} \left[\sum_{n\in \mathcal{S}_{\tau-1}} e^{i\theta_n}\ket{0}\ket{n} + \sum_{n\in \Pi^{2^{t-\tau}}(\mathcal{S}_{\tau-1})} e^{i(\theta_n+\varphi_\tau)}\ket{1}\ket{n}\right],
\end{equation}
\end{subequations}
where in the last equation $\mathcal{S}_{\tau-1}$ is a set containing $2^{\tau-1}$ distinct computational basis states of the $n$-qubit register. The phases $\theta_n$ and $\varphi_\tau$ depend on the measurement outcomes and on the phase gates, but their exact values are not crucial for the purpose of our study, and can thus be effectively thought as random.

Let $r$ be the period of the permutation $\Pi$, i.e., the smallest integer such that $\Pi^r = \mathds{1}$. Let us assume that it factorizes as $r=2^k \Tilde{r}_\mathrm{odd}$, where $k\geq 0$ and $\Tilde{r}_\mathrm{odd}$ is an odd integer. This implies that the period of $\Pi^{2^j}$ is equal to $\max\{\Tilde{r},\frac{r}{2^j}\}$. The superposition structure shown in Eq.~\eqref{ramp_wavefunction} holds as long as $\tau \leq \tau^* = \lceil \log_2 \Tilde{r}_\mathrm{odd}\rceil$. In detail, for $\tau$ strictly smaller than $\tau^*$, the sets $\mathcal{S}_{\tau-1}$ and $\Pi^{2^{t-\tau}}(\mathcal{S}_{\tau-1})$ are non-overlapping, because not enough computational basis states have been yet generated to cover a full permutation cycle of $\Pi^{2^{t-\tau}}$. For $\tau = \tau^*$, the equation is still valid, but the two sets overlap partially. 

\textbf{Intermediate times} -- For $\tau> \tau^*$, Eq.~\eqref{ramp_wavefunction} no longer applies. The sets $\mathcal{S}_{\tau-1}$ and $\Pi^{2^{t-\tau}}(\mathcal{S}_{\tau-1})$ appearing in Eq.~\eqref{ramp_wavefunction} coincide for $\tau \leq t-k$, and have size $\Tilde{r}_\mathrm{odd}$. As a consequence, the Hadamard gate placed before the measurements mixes the $\ket{0}$ and $\ket{1}$ sectors of the QFT qubit, creating interference and possibly making the superposition non-uniform. The state assumes the general form $\ket{\psi(\tau)} = \sum_{n\in \mathcal{S}} (a_{0,n}\ket{0} + a_{1,n}\ket{1})\ket{n} $, where $\mathcal{S}$ no longer depends on $\tau$. We refer to the window $\tau^*<\tau\leq t-k$ as the intermediate time regime, and we assume it in the following discussion.

Within this time window, the amplitudes of the superposition are repeatedly shuffled by the Hadamard gates, the phase gates, and the projective measurements applied at each step. Naively, one might expect that the result of this process is a Gaussian distribution for the real and imaginary parts of $a_{0,n}$ and $a_{1,n}$, following the intuition that the central limit theorem might apply. However, this is not the case. Let us analyze how amplitudes are recombined during a step of the algorithm. Let the input state of the $\tau$-th algorithmic step be given by $\ket{\psi_\mathrm{in}(\tau)} = \ket{0}\sum_{n\in \mathcal{S}}a_n(\tau)\ket{n}$. Throughout gates (i)-(iii), we obtain
\begin{equation}\label{psi_mid}
    \ket{\psi(\tau)} = \frac{1}{\sqrt{2}}\sum_{n\in \mathcal{S}} \left(a_n \ket{0} + e^{i\varphi_\tau} a_{\Pi^{-2^{(t-\tau)}}(n)}\ket{1}\right) \ket{n}.
\end{equation}
After applying the operations (iv) and (v), we have $\ket{\psi_\mathrm{in}(\tau+1)} = \frac{1}{\mathcal{N}_\tau}\ket{0}\sum_{n\in \mathcal{S}}\left(a_n(\tau) + (-1)^{m_\tau} e^{i\varphi_\tau} a_{\Pi^{-2^{(t-\tau)}}(n)}(\tau)\right)\ket{n}$, where $m_\tau = 0,1$ is the outcome and $\mathcal{N}_\tau$ is a normalization factor. If we introduce the array $\mathbf{a}(\tau) = (a_n)_{n \in \mathcal{S}}$, the previous equation defines the evolution map
\begin{equation}
    \mathbf{a}(\tau+1) = \frac{1}{\mathcal{N}_\tau}\left(\mathds{1}+(-1)^{m_\tau}e^{i \varphi_\tau}\Pi^{-2^{(t-\tau)}}\right) \mathbf{a}(\tau) \equiv \mathcal{M}(\tau)\mathbf{a}(\tau) ,
\end{equation}
with a slight abuse of notation on the permutation $\Pi$ for the sake of brevity.

It is now crucial to observe that the operators $\mathcal{M}(\tau)$ at different times commute. This allows us to diagonalize them and extract the long-time behavior. First, since we assumed $\tau\leq t-k$, all $\Pi^{-2^{(t-\tau)}}$ are powers of $\Tilde{\Pi} = \Pi^{2^k}$, which is an irreducible permutation with period $\Tilde{r}_\mathrm{odd}$. We diagonalize $\Tilde{\Pi} = U S U^\dagger$, where $S=\mathrm{diag}\{1,e^{2 \pi i/\Tilde{r}_\mathrm{odd}},\dots, e^{2 \pi i (\Tilde{r}_\mathrm{odd}-1)/\Tilde{r}_\mathrm{odd}}\}$ and $U_{\mu,\nu} = e^{2\pi i b(\mu) \nu /\Tilde{r}_\mathrm{odd}}/\sqrt{\Tilde{r}_\mathrm{odd}}$, with $b(\mu)\in \{0,\dots, \Tilde{r}_\mathrm{odd}-1\}$. It follows that
\begin{equation}\label{diagonal_map}
    \mathrm{a}_\mu(\tau+n) = \frac{1}{\prod_{j=1}^n \mathcal{N}_{\tau+j}} \sum_{\mu,\alpha,\nu=1}^{\Tilde{r}_\mathrm{odd}}U_{\mu,\alpha}\prod_{j=1}^n\left(1+e^{i\theta_{j,\alpha}}\right) U^*_{\nu,\alpha} \mathrm{a}_\nu(\tau),
\end{equation}
where we defined $\theta_{j,\alpha} = \pi m_{\tau+j} + \varphi_{\tau+j} + 2^{t-\tau-k+1}\pi \alpha/\Tilde{r}_\mathrm{odd}$.

Each value of $\alpha$ contributes differently to Eq.~\eqref{diagonal_map}, and a single one will eventually dominate for large $n$. To see this, we may evaluate
\begin{equation}\label{log_sum}
\log\left|\prod_{j=1}^n\left(1+e^{i\theta_{j,\alpha}}\right)\right| = \sum_{j=1}^n \log\left|2 \cos(\theta_{j,\alpha}/2)\right|.
\end{equation}
The angles $\theta_{j,\alpha}$ hold no particular structure, and can thus be regarded as pseudo-random variables. From the central limit theorem, for $n\gg 1$ Eq.~\eqref{log_sum} behaves as a Gaussian random variable with variance $\sim n$. As a consequence, the typical difference in magnitude between the contributions of different values of $\alpha$ in Eq.~\eqref{diagonal_map} will be exponentially large in $\sqrt{n}$, thus implying that a single $\Bar{\alpha}$ dominates the sum at long times. This allows us to approximate
\begin{equation}\label{diagonal_map}
    \mathrm{a}_\mu(\tau+n) \approx \left(\prod_{j=1}^n\frac{1+e^{i\theta_{j,\Bar{\alpha}}}}{\mathcal{N}_{\tau+j}}\right) \left(\sum_{\nu=1}^{\Tilde{r}_\mathrm{odd}}U^*_{\nu,\Bar{\alpha}} \mathrm{a}_\nu(\tau) \right) U_{\mu,\Bar{\alpha}}.
\end{equation}
Notice that the dependence on the initial amplitudes is shared equally among all values of $\mu$. Thus, accounting for the correct normalization, the equation simplifies drastically to $\mathrm{a}_\mu(\tau+n) = U_{\mu,\Bar{\alpha}} = e^{2\pi i b(\mu)\Bar{\alpha}  /\Tilde{r}_\mathrm{odd}}/\sqrt{\Tilde{r}_\mathrm{odd}}$. This implies that during the intermediate time regime the amplitudes quickly converge to a uniform distribution around the circle of radius $1/\sqrt{\Tilde{r}_\mathrm{odd}}$.

From this result, it might seem as if the superposition structure is approximately the same in the early and intermediate time regimes of the algorithm. There is, however, a crucial difference. From Eq.~\eqref{psi_mid}, we see that the amplitudes of the QFT-qubit sector $\ket{1}$ are given by $e^{i \varphi_\tau} \Pi^{-2^{t-\tau}} \mathbf{a}(\tau)$. Since $\mathbf{a}(\tau)$ converges to an eigenvector of the permutation, the expression simplifies to $e^{i (\varphi_\tau-2^{t-\tau-k+1}\pi \Bar{\alpha}/\Tilde{r}_\mathrm{odd})} \mathbf{a}(\tau)$, thus reducing Eq.~\eqref{psi_mid} to
\begin{equation}\label{plateau_wavefunction}
    \ket{\psi(\tau)} = \frac{\ket{0}+e^{i\phi }\ket{1}}{\sqrt{2}}\sum_{n \in \mathcal{S}} \frac{e^{i\theta_n}} {\sqrt{\Tilde{r}_\mathrm{odd}}}\ket{n}.
\end{equation}
The QFT qubit disentangles completely from the others. The phases $\theta_n$ correspond to the entries of $U_{\mu,\Bar{\alpha}}$ and can be considered effectively random. 

We checked numerically the validity of Eq.~\eqref{plateau_wavefunction}. In the process, we observed that, at sufficiently long times, the relative phase $\phi$ is not random, but becomes either $0$ or $\pi$. While we have no analytic proof of this result, it is fully reasonable. The probability distribution of the measurement outcomes of Shor's algorithm is highly peaked around some specific bitstrings, which contain information on the period $r$. In order for this to be possible, the probabilities of getting $0$ and $1$ during a single measurement must be strongly biased towards one of the two, in such a way to select only a restricted set of sequences of outcomes. This corresponds precisely to the condition $\phi \approx 0, \pi$.

\textbf{Late times} -- At last, if the period $r$ is even, the structure of the state changes again. During the last $k$ steps of the algorithm, the permutation $\Pi^{2^{(t-\tau)}}$ starts again acting non-trivially on the set $\mathcal{S}$, thus generating new computational basis states in the superposition. This immediately implies that no further interference between the $\ket{0}$ and $\ket{1}$ sectors of the QFT qubit can occur: the superposition starts doubling in size at each step, analogously to what happens in the early time regime. As a consequence, for $\tau > t-k$, the state assumes the form of Eq.~\eqref{ramp_wavefunction} once again, where $\mathcal{S}_{t-k}$ is the set $\mathcal{S}$ of size $\Tilde{r}_\mathrm{odd}$ of the intermediate time regime.

\section{Analytic calculation of the SRE}

Given the knowledge of the superposition structure of the state throughout the algorithm, we can now proceed to evaluate the second stabilizer Renyi entropy explicitly. We use the formalism of the fast Walsh-Hadamard transform introduced in Ref.~\cite{huang2026}. Let us assume that the state is in the generic form $\ket{\psi} = \sum_{\mathbf{m}\in \mathcal{D}}a_\mathbf{m}\ket{\mathbf{m}}$, where $\mathcal{D}$ is the set of bitstrings $\mathbf{m}$ corresponding to computational basis states with non-zero amplitudes in the superposition. For both the early and late time regimes, the state is given by Eq.~\eqref{ramp_wavefunction}. In the intermediate window, as argued previously, we have Eq.~\eqref{plateau_wavefunction}, and the QFT qubit factorizes in a stabilizer state corresponding to $\phi=0,\pi$. As a consequence, that qubit can be effectively discarded for the calculation of magic, and we may just assume that $\ket{\psi}$ is the state of the other $n = L-1$ qubits, so that $\mathcal{D}$ corresponds to the set $\mathcal{S}$ of size $D=\Tilde{r}_\mathrm{odd}$. Following our previous discussion, in both cases amplitudes are modeled in the form of $a_\mathbf{m} = e^{i\theta_\mathbf{m}}/\sqrt{D}$, where $\theta_\mathbf{m}$ are independent random angles uniformly distributed in $[0,2\pi)$. Normalization sets the value $D = \dim \mathcal{D}$. 

In order to evaluate the SRE of Eq.~\eqref{SRE}, we parameterize Pauli strings as
\begin{equation}
    P \equiv P_{\mathbf{z},\mathbf{x}}= i^{-\mathbf{z}\cdot \mathbf{x}}\left(\prod_{j=1}^L Z_j^{z_j}\right)\left(\prod_{j=1}^L X_j^{x_j}\right),
\end{equation}
where $\alpha\in\{0,1,2,3\}$, $\mathbf{x},\mathbf{z}\in\{0,1\}^L$. We thus have $\langle P_{\mathbf{z},\mathbf{x}} \rangle = i^{-\mathbf{z}\cdot \mathbf{x}} \sum_{\mathbf{m}\in \mathcal{D}} a^*_\mathbf{m} a_{\mathbf{m}\oplus \mathbf{x}} (-1)^{\mathbf{m}\cdot\mathbf{z}} \mathds{1}(\mathbf{m}\oplus \mathbf{x} \in \mathcal{D})$, where we introduced the indicator function $\mathds{1}$ that returns $1$ if its Boolean argument is True, and $0$ otherwise. From this expression, we obtain
\begin{equation}\label{pauli_sum}
    \frac{1}{2^L}\sum_P \langle P \rangle^4 = \sum_{\mathbf{m},\mathbf{n},\mathbf{p},\mathbf{q}\in \mathcal{D}} \delta_{\mathbf{m}\oplus\mathbf{n}\oplus\mathbf{p}\oplus\mathbf{q},\mathbf{0}} \sum_\mathbf{x} a^*_\mathbf{m}a_{\mathbf{m\oplus x}} a^*_\mathbf{n}a_{\mathbf{n\oplus x}} a^*_\mathbf{p}a_{\mathbf{p\oplus x}} a^*_\mathbf{q}a_{\mathbf{q\oplus x}} \mathds{1}(\{\mathbf{m},\mathbf{n},\mathbf{p},\mathbf{q}\}\oplus \mathbf{x} \subset \mathcal{D}).
\end{equation}

Only bitstring quadruplets satisfying $\mathbf{m}\oplus\mathbf{n}\oplus\mathbf{p}\oplus\mathbf{q} = \mathbf{0}$ contribute to the sum of Eq.~\eqref{pauli_sum}. There are three ways to guarantee this condition:
\begin{enumerate}
    \item All bitstrings of the quadruplet are equal. In this case, we get a contribution $\sum_{\mathbf{m}\in \mathcal{D}} \sum_\mathbf{x}(a^*_\mathbf{m} a_{\mathbf{m\oplus \mathbf{x}}})^4 \mathds{1}(\mathbf{m}\oplus \mathbf{x}\in \mathcal{D})$. For $D\gg 1$, the sum can be approximated by a statistical average over the distribution of amplitudes, which we denote by $\mathds{E}[\dots]$. Since $\mathds{E}[(a^*_\mathbf{m} a_{\mathbf{m\oplus \mathbf{x}}})^4]=\mathds{E}[(a^*_\mathbf{m} )^4] \mathds{E}[(a_{\mathbf{m\oplus \mathbf{x}}})^4]=0$ for $\mathbf{x}\neq \mathbf{0}$, only $\mathbf{x} = \mathbf{0}$ makes the sum non-zero, yielding a final contribution $1/D^3$ to the Pauli sum.
    \item The bitstrings of the quadruplets are equal in pairs. In this case, the contribution to Eq.~\eqref{pauli_sum} is equal to $3\sum_{\mathbf{m}\neq\mathbf{n}\in \mathcal{D}}\sum_\mathbf{x}(a^*_\mathbf{m} a_{\mathbf{m\oplus \mathbf{x}}}a^*_\mathbf{n} a_{\mathbf{n\oplus \mathbf{x}}})^2 \mathds{1}(\{\mathbf{m},\mathbf{n}\}\oplus\mathbf{x}\subset \mathcal{D})$, where the factor $3$ counts the number of distinct ways to divide the four bitstrings in distinct pairs. As done previously, the sum can be approximated by a statistical average. Both $\mathbf{x} = \mathbf{0}$ and $\mathbf{x} = \mathbf{m}\oplus\mathbf{p}$ produce non-zero averages, thus giving a total contribution of $6(D-1)/D^3$.
    \item All bitstrings of the quadruplet are distinct. For each choice of the four bitstrings satisfying $\mathbf{m}\oplus\mathbf{n}\oplus\mathbf{p}\oplus\mathbf{q}=\mathbf{0}$, the distinct values of $\mathbf{x}$ that give a non-zero contribution are $\mathbf{x}=\mathbf{0}$, $\mathbf{x}=\mathbf{m}\oplus\mathbf{n}$, $\mathbf{x}=\mathbf{m}\oplus\mathbf{p}$, and $\mathbf{x}=\mathbf{m}\oplus\mathbf{q}$. As a consequence, the total contribution of these terms to the Pauli sum is $4 \Lambda /D^4$, where we define
    \begin{equation}\label{lambda}
    \Lambda = \sum_{\mathbf{m}\neq\mathbf{n}\neq\mathbf{p}\neq\mathbf{q}\in \mathcal{D}} \delta_{\mathbf{m}\oplus\mathbf{n}\oplus\mathbf{p}\oplus\mathbf{q},\mathbf{0}}.
    \end{equation}
    Differently from the previous cases, this term is not uniquely determined by the size of $\mathcal{D}$. The number $\Lambda$ cannot be evaluated explicitly without additional assumptions.
\end{enumerate}

Combining these calculations, we obtain the final form
\begin{equation}\label{sre_closed}
    M_2 = \begin{cases}
    0                                                   & \text{if} \,\,\, D=2,\\
    4 \log D - \log\left( 4\Lambda + 6 D^2 - 5 D\right) & \text{otherwise}.
    \end{cases}
\end{equation}
Here, we distinguished the special case of $D=2$, which behaves differently from the other values and can be proven to have vanishing magic. In general, we always find $D=2$ for $\tau=1$, in which case the state is given exactly by Eq.~\eqref{psi_1}: this is clearly a stabilizer state, and thus its magic is zero. In addition, we observe that $D=2$ at all times $\tau$ when the period of the permutation $\Pi$ is exactly $r=2$. In this case, since $\Pi^{2^j}$ is the identity for any $j>0$, it can be easily checked that the state remains a stabilizer throughout the full algorithm.

In order to compute $M_2$ explicitly as a function of the discrete time $\tau$, we need to evaluate how $\mathcal{D}$ grows. Following the discussion of Sec.~\ref{s:wavefunction}, if the period of the permutation $\Pi$ can be factored as $r = 2^k\Tilde{r}_\mathrm{odd}$, we have
\begin{equation}\label{size_closed}
    D = \begin{cases}
        2^\tau                 & \text{for} \,\,\,\tau\leq \tau^*,\\
        \Tilde{r}_\mathrm{odd} & \text{for} \,\,\,\tau^*<\tau\leq t-k,\\
        r/2^{t-\tau}           & \text{for}\,\,\,t-k<\tau.
    \end{cases}
\end{equation}

\textbf{Assumption of random bitstrings} -- The combination of Eqs.~\eqref{sre_closed} and~\eqref{size_closed} provides a semi-analytic prediction for the growth of the SRE. This is not, however, a closed formula, as $\Lambda$ needs to be evaluated numerically. This calculation is expensive, as the size of $\mathcal{D}$ can grow exponentially in the number of qubits. Nevertheless, $\Lambda$ can be approximated under the reasonable assumption that the bitstrings contained in $\mathcal{D}$ can be regarded as random, sampled uniformly from the computational basis.

Suppose we pick the first three bitstrings $\mathbf{m}$, $\mathbf{n}$, and $\mathbf{p}$ from $\mathcal{D}$. There are $D(D-1)(D-2)$ ways of performing these choices. The probability that the specific $\mathbf{q}=\mathbf{m}\oplus \mathbf{n} \oplus \mathbf{p}$ is also in the set is equal to $(D-3)/2^L$ for early and late times, and to $(D-3)/2^{L-1}$ for intermediate times, because in the latter case we are not considering the QFT qubit. This yields
\begin{equation}\label{lambda_closed}
    \Lambda = \begin{cases}
        \frac{D(D-1)(D-2)(D-3)}{2^{L-1}} \qquad \text{for}\,\,\,\lceil\log_2r\rceil-k<\tau\leq t-k,\\
        \frac{D(D-1)(D-2)(D-3)}{2^L}\qquad \text{otherwise}. 
        \end{cases}
\end{equation}
This estimate is valid asymptotically for $L\to\infty$ and $D\gg 1$.

\end{document}